\pdfoutput=1
\documentclass[conference,onecolumn]{IEEEtran}
\usepackage{acronym}

\acrodef{fmcw}[FMCW]{frequency modulated continuous wave}
\acrodef{mmic}[MMIC]{monolithic microwave integrated circuit}
\acrodef{adc}[ADC]{analog-to-digital converter}
\acrodef{dac}[DAC]{digital-to-analog converter}
\acrodef{sfdr}[SFDR]{spurious-free dynamic range}
\acrodef{seir}[SEIR]{stimulus error identification and removal}
\acrodef{sh}[S\&H]{sample and hold}
\acrodef{hee}[HEC]{homogeneity enforced calibration}
\acrodef{mdac}[MDAC]{multiplying DAC}
\acrodef{enob}[ENOB]{effective number of Bits}
\acrodef{inl}[INL]{integral nonlinearity}
\acrodef{lms}[LMS]{least mean squares}
\acrodef{pn}[PN]{phase noise}
\acrodef{dpn}[DPN]{decorrelated phase noise}
\acrodef{if}[IF]{intermediate frequency}
\acrodef{dut}[DUT]{device under test}
\acrodef{sg}[SG]{signal generator}
\acrodef{mmse}[MMSE]{minimum mean squared error}
\acrodef{mse}[MSE]{mean squared error}
\acrodef{sa}[SA]{system application}
\acrodef{sgd}[SGD]{stochastic gradient descent}
\acrodef{pdf}[PDF]{probability density function}
\acrodef{als}[ALS]{approximate least squares}

\usepackage[nospace]{cite}
\ifCLASSINFOpdf
\else
\fi
\usepackage[pdftex]{graphicx}

\usepackage{pgfplots}
\pgfplotsset{compat=newest}
\pgfplotsset{plot coordinates/math parser=false}
\usepackage{tikz}
\tikzset{>=latex}
\usetikzlibrary{calc,arrows.meta,positioning}

\usepackage{xcolor}
\definecolor{lightorange}{rgb}{1,0.66,0.16}
\definecolor{isp_blue}{RGB}{14,98,167}
\definecolor{lightgreen}{rgb}{0,0.8,0}

\usepackage[cmex10]{amsmath}
\usepackage{amssymb}
\usepackage[font=footnotesize]{subfig}
\newcommand{\E}[2]{\ensuremath{{E}_{#2}\left[#1\right]}}

\usepackage{blindtext}
\usepackage{todonotes}
\usepackage{romannum}
\usepackage{endnotes}

\begin{document}
%
\title{Homogeneity Enforced Calibration of Stage Nonidealities for Pipelined ADCs}

\author{Matthias Wagner$^1$, Oliver Lang$^1$, Thomas Bauernfeind$^2$ and Mario Huemer$^1$ \vspace{0.1cm} \\
        \normalsize $^1$ Institute of Signal Processing, Johannes Kepler University, Linz\\
        $^2$ Infineon Technologies Linz GmbH \& Co. KG}


%


\maketitle

\begin{abstract}
	Pipelined \acp{adc} are fundamental components of various signal processing systems requiring high sampling rates and a high linearity. Over the past years, calibration techniques have been intensively investigated to increase the linearity. \\
	In this work, we propose an equalization-based calibration technique which does not require knowledge of the \ac{adc} input signal for calibration. For that, a test signal and a scaled version of it are fed into the \ac{adc} sequentially, while only the corresponding output samples are used for calibration. Several test signal sources are possible, such as a \ac{sg} or the \ac{sa} itself. For the latter case, the presented method corresponds to a background calibration technique. Thus, slowly changing errors are tracked and calibrated continuously.  Because of the low computational complexity of the calibration technique, it is suitable for an on-chip implementation. Ultimately, this work contains an analysis of the stability and convergence behavior as well as simulation results.
\end{abstract}


\section{Introduction}
Many state-of-the-art signal processing systems require \acfp{adc} with a high sampling rate, a high dynamic range, and a high linearity. Pipelined \acp{adc} are a popular choice for such challenging applications. As discussed in \cite{14_ADC_HighSpeedTimeInterleavedADCs}, the time-interleaved architecture of pipelined \acp{adc} enables high sampling frequencies, albeit it introduces additional error sources. Besides random errors, i.\,e. thermal and quantization noise, \ac{dac} or gain mismatches cause systematic errors. To maintain a high linearity, calibration techniques for these systematic errors become inevitable and were heavily investigated throughout the past decades \cite{19_ADC_CalibrationAndDynamicMatchinInDataConverters1,20_ADC_CalibrationAndDynamicMatchinInDataConverters2}. To reduce power consumption and relax the analog design, calibration techniques are preferably carried out in the digital domain \cite{12_ADC_ASurveyOnDigitalBackgroundCalibrationOfADCs}.
These techniques may be grouped in equalization-\nolinebreak, histogram-, and correlation-based approaches. A crucial part of calibration techniques is the identification of the \ac{adc} nonidealities \cite{12_ADC_ASurveyOnDigitalBackgroundCalibrationOfADCs}.  Typically, this requires either long test times \cite{4_ADC_DigitalBackgroundCalibrationWithHistogramOfDecisionPointsInPipelinedADCs,5_ADC_ADigitalBackgroundCalibrationSchemeForPipelinedADCsUsingMultipleCorrelationEstimation,7_ADC_StatisticsBasedDigitalBackgroundCalibrationOfResidueAmplifierNonlinearityInPipelinedADCs},
exceedingly linear test \acfp{sg} \cite{21_ADC_AlgorithmForDramaticallyImprovedEfficiencyInADCLinearityTest,22_ADC_LinearityTestingIssuesOfAnalogToDigitalConverters}, or additional hardware components \cite{1_ADC_LeasMeanSquareAdaptiveDigitalBackgroundCalibrationOfPipelinedAnalogToDigitalConverters,3_ADC_AnAdaptiveDigitalBackgroundCalibrationTechniqueUsingVariableStepSizeLMSForPipelinedADC}.
\\
Much effort was spent to overcome the need of a highly linear test \ac{sg} \cite{13_ADC_ADCTestMethodsUsingAnImpureStimulus}. In  \cite{15_ADC_AccurateTestingOfAnalogTotDigitalConvertersUsingLowLinearitySignalsWithStimulusErrorIdentification}
a histogram-based method is introduced, which relaxes the linearity requirements of the test \ac{sg} heavily. This is achieved by injecting a test signal twice to the \ac{adc}, whereas for the second time, it is shifted by a constant voltage-offset. Therewith, the nonlinear test signal is identified and it is possible to subsequently estimate the \ac{adc} nonideality. In literature, this method is referred to as the \ac{seir} approach. Improvements can be found in \cite{18_ADC_HighPrecisionADCTestingWithRelaxedReferenceVoltageStationarity,16_ADC_ARobustAlgorithmToIdentifyTheTestStimulusInHistogramBasedADCTesting}, which address enhanced robustness concerning reference voltage stationarity and a reduced hardware overhead, respectively. The identification of the \ac{adc} nonidealities in \cite{11_ADC_USERSMILEUltrafastStimulusErrorRemovalAndSegmentedModelIdentificationOfLinearityErrorsForADCBuiltInSelfTest} is based on the same principle as the \ac{seir} approach. However, by using a segmented nonlinearity model it significantly reduces the required test time.  \\
As the \ac{adc} under calibration is nonlinear, it violates the linearity conditions. Specifically, a nonlinear system $f(x):\mathbb{R}\rightarrow\mathbb{R}$ violates at least one of the following two conditions:
\begin{IEEEeqnarray}{rCl"s}
	f(\alpha x) & = & \alpha f(x) &\ldots\text{ homogeneity}
	\label{eq_hom}\\
	f(x+y) & = & f(x) + f(y) &\ldots\text{ additivity}
	\label{eq_add}
\end{IEEEeqnarray}
with $x$, $y$ as the system inputs and an arbitrary constant $\alpha \in\mathbb{R}$. A closer look at the principle of \cite{11_ADC_USERSMILEUltrafastStimulusErrorRemovalAndSegmentedModelIdentificationOfLinearityErrorsForADCBuiltInSelfTest} shows that in this sense, it relies on a weak form of additivity, where the constant voltage-offset replaces the second input signal $y$ in (\ref{eq_add}). \\
In this work, we propose a calibration technique that focuses on the homogeneity condition in (\ref{eq_hom}).  The \ac{adc} calibration is achieved by injecting a test signal twice into the \ac{adc}, whereas for the second time, it is scaled by a constant factor $\alpha$.  From here on the calibration technique will be referred to as \ac{hee} approach. By scaling the input signal with $\alpha$, the \ac{hee} approach replaces the necessity for a highly constant voltage-offset. The scaling, however, can easily be implemented using a voltage divider. Along with a low complex post correction model, the \ac{hee} approach is suitable for off- as well as on-chip calibration. Nonetheless, injecting a test signal twice prevents an implementation without the interruption of the sampling task of the \ac{adc}. Therefore, this work additionally introduces an extension of the \ac{hee} approach that allows for background calibration. It will be shown that the \ac{hee} approach exhibits fast convergence times and a low computational complexity, while requiring only a few additional hardware components.\\
The rest of this paper is organized as follows. In Section\,\ref{seq_systemModel}, a detailed derivation of the \ac{adc} output signal including its nonidealities is presented. A post correction model is discussed in Section\,\ref{sec_postcorr}. The \ac{hee} approach is derived in Section\,\ref{sec_hec} including analysis of its convergence and stability behavior. The findings are numerically verified in Section\,\ref{sec_SIM}. Finally, Section\,\ref{sec_con} concludes this work. 

\section{System Model}
\label{seq_systemModel}
\begin{figure}[!t]
	\centering
	\includegraphics{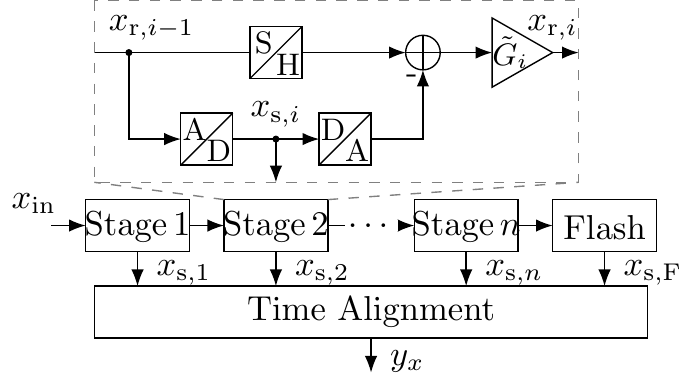}
	\caption{Block diagram of a pipelined \ac{adc} including the stage building blocks.}
	\label{fig_pipelinedADC}
\end{figure}

This section provides an introduction to the \ac{adc} model employed in this work. Furthermore, the impact of \ac{dac} and gain mismatches caused, e.\,g., by component variation, aging- or temperature effects, on the transition function of a pipelined \ac{adc} is derived in detail.\\
Typically, pipelined \acp{adc} consist of $n$ equally built stages followed by a final flash \ac{adc} stage. As Fig.\,\ref{fig_pipelinedADC} shows, each stage involves a \ac{sh} block, a stage \ac{adc}, a stage \ac{dac}, a subtracter, and a residue amplifier. Conventional architectures combine the \ac{dac}, the subtracter, and the amplifier in a switched capacitor circuit, referred to as a multiplying \ac{dac} entity. Because of the \ac{sh} blocks, each stage simultaneously resolves a stage output with respect to consecutive input voltages. These stage outputs $x_{\text{s},i}$, $i = 1,\ldots,n$, are then time aligned and combined to form the overall \ac{adc} output $y_x$, where the index $i$ refers to the stage under consideration. By neglecting the time alignment to increase readability, the output can be expressed as   
\begin{equation}
	y_x = x_{\text{s},1}+\frac{x_{\text{s},2}}{G_1}+\frac{x_{\text{s},3}}{G_1G_2}+\ldots+\frac{x_{\text{s},\text{F}}}{\prod_{i=1}^{n}G_i},
	\label{eq_adcoutput1}
\end{equation}
with the ideal stage gains $G_i$ and $x_{\text{s,F}}$ denoting the output of the final flash \ac{adc}. In the presence of \ac{dac} and gain mismatches, the real stage outputs do not match the ideal ones. Thus, the output signal in (\ref{eq_adcoutput1}) is distorted. 
According to Fig.\,\ref{fig_pipelinedADC}, the residue signal of each stage may be written as
\begin{equation}
	x_{\text{r},i} = \tilde{G}_i\left(x_{\text{r},i-1}-x_{\text{s},i}-e_{\text{s},i}^{\text{DA}}(x_{\text{s},i})\right),
	\label{eq_residuesig1}
\end{equation}
where $\tilde{G}_i$ denotes nonideal stage gains and $e_{\text{s},i}^{\text{DA}}(x_{\text{s},i})$ represents errors caused by \ac{dac} mismatches. The argument of $e_{\text{s},i}^{\text{DA}}(x_{\text{s},i})$ should point out that each stage output level can result in a different \ac{dac} error. By introducing the quantization error $e_{\text{s},i}^{\text{q}}$, the stage output can be expressed as
\begin{equation}
	x_{\text{s},i} = x_{\text{r},i-1}- e_{\text{s},i}^\text{q}.
	\label{eq_quanterr}
\end{equation}
Inserting (\ref{eq_quanterr}) into (\ref{eq_residuesig1}) yields the transformed residue signal
\begin{equation}
	x_{\text{r},i} = \tilde{G}_i\left( e_{\text{s},i}^\text{q}-e_{\text{s},i}^{\text{DA}}(x_{\text{s},i})\right),
	\label{eq_residuesig2}
\end{equation}
which depends only on nonideal stage gains, the stage quantization signal and the \ac{dac} errors.
With (\ref{eq_quanterr}) and (\ref{eq_residuesig2})  the \ac{adc} output signal in (\ref{eq_adcoutput1}) may be written as
\begin{equation}
	\begin{split}
		y_x =\text{\space}& x_{\text{in}} + e_{\text{s},1}^\text{q}\left(\frac{\tilde{G}_1}{G_1}-1\right)-e_{\text{s},1}^{\text{DA}}\frac{\tilde{G}_1}{G_1}+ e_{\text{s},2}^\text{q}\frac{1}{G_1}\left(\frac{\tilde{G}_2}{G_2}-1\right)-e_{\text{s},2}^{\text{DA}}\frac{\tilde{G}_2}{G_1G_2}+\ldots+\frac{ e_{\text{s},\text{F}}^\text{q}}{\prod_{i=1}^{n}G_i}, 
	\end{split}
	\label{eq_adcoutput2}
\end{equation}
where $ e_{\text{s},\text{F}}^\text{q}$ denotes the quantization error of the final flash \ac{adc}. 
\begin{figure}[!t]
	\centering
	\includegraphics{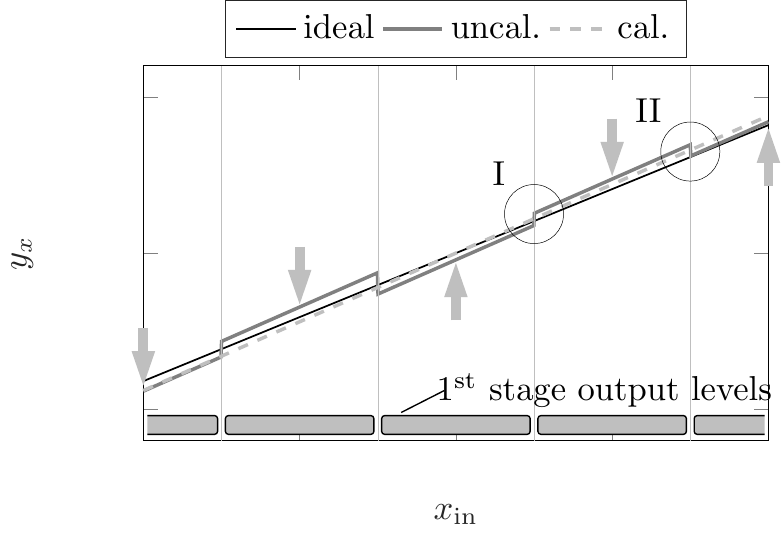}
	\caption{Transition function affected by errors in the first stage.}
	\label{fig_TF1}
\end{figure}
For the sake of readability, the argument of the \ac{dac} error $e_{\text{s},i}^{\text{DA}}(x_{\text{s},i})$ is omitted as it is clear from the context.
With the relative gain errors $\zeta_i$, the nonideal residue gain may be written as $\tilde{G}_i = G_i\left(1+\zeta_i\right)$. Substituting the nonideal gain into (\ref{eq_adcoutput2}) yields the final expression for the output signal as
\begin{equation}
	\begin{split}
		y_x =\text{\space}& x_\text{in}+\zeta_1 e_{\text{s},1}^\text{q}-\left(1+\zeta_1\right)e_{\text{s},1}^{\text{DA}}+\frac{\zeta_2}{G_1} e_{\text{s},2}^\text{q}-\frac{1}{G_1}\left(1+\zeta_2\right)e_{\text{s},2}^{\text{DA}}+\ldots+\frac{\zeta_n}{\prod_{i=1}^{n-1}G_i} e_{\text{s},n}^\text{q}-\frac{1}{\prod_{i=1}^{n-1}G_i}\left(1+\zeta_n\right)e_{\text{s},n}^{\text{DA}}+\frac{ e_{\text{s},\text{F}}^\text{q}}{\prod_{i=1}^{n}G_i}.
	\end{split}
	\label{eq_INL}
\end{equation}
This equation shows, that nonidealities in less significant stages marginally affect the overall \ac{adc} output because they are weighted with the inverse product of all previous stage gains. Consequently, sufficient calibration performances for many applications might be achieved by considering only the most significant stages \cite{2_ADC_EqualizationBasedDigitalBackgroundCalibrationTechniqueForPipelinedADCs,3_ADC_AnAdaptiveDigitalBackgroundCalibrationTechniqueUsingVariableStepSizeLMSForPipelinedADC}. Additionally, (\ref{eq_INL}) shows that $\zeta_i \neq 0$ distorts the transition function by a remaining stage quantization error. \ac{dac} errors, on the other hand, add an offset to the corresponding stage output levels. This is illustrated in Fig.\,\ref{fig_TF1}, which shows an exemplary transition function of a pipelined \ac{adc} affected by errors in the first stage. Additionally, Fig.\,\ref{fig_TF1} highlights two types of discontinuities, i.\,e. missing codes (\Romannum{1}), and nonmonotonous errors (\Romannum{2}). Because of these discontinuities, the transition function is not invertible, such that applying the inverse system for calibration is not feasible. Nonetheless, as shown in \cite{9_ADC_BlackBoxCalibrationForADCsWithHardNonlinearErrorsUsingANovelINLBasedAdditiveCode}, this issue may be coped by using the stage outputs separately for calibration.
\section{Post Correction Model}
\label{sec_postcorr}
The post correction model employed in this work corrects the discontinuities by adding corrective offsets to the corresponding stage output levels. In Fig.\,\ref{fig_TF1}, the output levels of the first stage are schematically illustrated. The added offsets serve as parameters to be identified for the calibration and are summarized in a parameter vector $\boldsymbol{\theta}$. Let $p$ denote the total number of stage output levels of all stages used for calibration. Storing all offsets would usually require a vector of length $p$. However, there are some dependencies between the offsets of consecutive stages, by means that one parameter in the $i$-th stage can be represented by an offset of all parameters in the $\left(i+1\right)$-th stage. Eliminating these dependencies by excluding the redundant entries of $\boldsymbol{\theta}$ results in a parameter vector of length $p-\left(q-1\right)$, with $q$ denoting the number of stages considered for calibration\footnote{In this work, always the parameter corresponding to the lowest stage output level is excluded.}. The entries of $\boldsymbol{\theta}$ that contribute to a particular \ac{adc} output are isolated by the selection vector $\mathbf{h}_{x}$. To do so, the corresponding entries of $\mathbf{h}_x$ are set to 1 and 0 for active and inactive output levels, respectively. However, gain errors do not only affect one dedicated stage output level but all corresponding stage output levels similarly. It can be shown that the employed post correction model is more effective if one parameter of each stage affects not only one stage output level but all output levels of the corresponding stage. This may be achieved by replacing $q$ entries of $\mathbf{h}_x$, which are related to different stages\footnote{Every output level providing $x_{\text{s},i} \neq 0$ may be utilized for the replacement. In this work, the highest stage output level of each considered stage is used.} with $z_j = G_{j-1}z_{j-1}+x_{\text{s},j}$, where $z_1 = x_{\text{s},1}$ and $j =2\ldots q$. For example, consider a two stage \ac{adc} with three output levels in each stage, as well as an input signal $x_\text{in}$ such that the middle output level of the first stage and the highest output level of the second stage is active. This yields the selection vector $\mathbf{h}_x^T = \begin{bmatrix}
	1&x_{\text{s},1}&0&0&G_1x_{\text{s},1}+x_{\text{s},2}
\end{bmatrix}$, with $\left(\cdot\right)^T$ denoting transposition. Finally, the post-corrected output signal can be written as
\begin{equation}
	y_x^\text{c} =y_x+\mathbf{h}_{x}^T\boldsymbol{\theta}.
	\label{eq_corrmod}
\end{equation}
The impact of this post correction model to the transition function is illustrated in Fig.\,\ref{fig_TF1}. Therein the arrows indicate the offsets applied for calibration, and the dashed line represents the corrected transition function. As can be seen, the post correction model only eliminates the discontinuities, while an overall gain error remains. This overall gain error, however, does not affect the \ac{adc} linearity and may be eliminated in an additional processing step. 
\section{Homogeneity Enforced Calibration}
\label{sec_hec}
\begin{figure}[!t]
	\centering
	\includegraphics{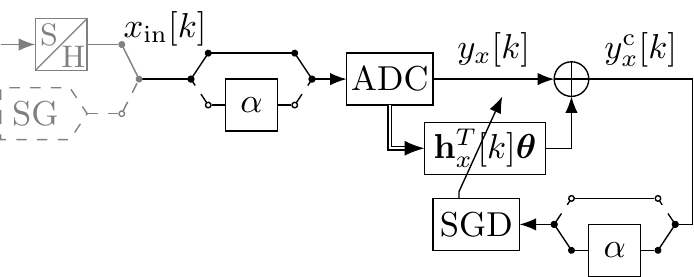}
	\caption{Block diagram of the \ac{hee} approach including different test signal sources.}
	\label{fig_calibration1}
\end{figure}

The basic structure of the \ac{hee} approach is shown in Fig.\,\ref{fig_calibration1}. There, the first step is to feed an input sample $x_\text{in}[k]$ into the \ac{adc}, where $k$ is the sample index. This sample may originate from an \ac{sg} or from the \acf{sa} itself stored in an \ac{sh} block. The corresponding output sample is denoted as $y_x[k]$. In the second step, the same input sample is scaled by a factor $\alpha$, and subsequently fed into the \ac{adc}. The resulting \ac{adc} output sample is denoted as $y_{\alpha x}[k]$. Since the \ac{adc} violates the homogeneity condition, the output $y_{\alpha x}[k]$ is in general not equal to $\alpha y_x[k]$. Applying the post correction in (\ref{eq_corrmod}) on both values yields the inequality 
\begin{equation}
	y_{\alpha x}[k]+\mathbf{h}_{\alpha x}^T[k]\boldsymbol{\theta}\neq \alpha \left(y_x[k]+\mathbf{h}_x^T[k]\boldsymbol{\theta}\right).
\end{equation}
The squared difference of these two terms is used as a cost-function according to 
\begin{align}
	J[k] = \left(y_{\alpha x}[k]+\mathbf{h}_{\alpha x}^T[k]\boldsymbol{\theta}-\alpha \left(y_{x}[k]+\mathbf{h}_{x}^T[k]\boldsymbol{\theta}\right)\right)^2.
	\label{eq_costfun0} 
\end{align}
A standard approach to identify the parameters $\boldsymbol{\theta}$ in the equation above is the method of least squares \cite{kay_estimation_theory}. This approach features a high computational complexity, furthermore, an appropriate number of output samples must be stored. Although the computational complexity may be relaxed with the \ac{als} algorithm \cite{Kaczmarz}, still storing an appropriate number of output samples is necessary. This drawback is overcome by adaptive algorithms, such as the \ac{hee} approach derived in the following. For simplification, $\Delta y[k] = y_{\alpha x}[k]-\alpha y_x[k]$ and $\Delta \mathbf{h}^T[k] = \mathbf{h}_{\alpha x}^T[k]-\alpha \mathbf{h}_{x}^T[k]$ are inserted in (\ref{eq_costfun0}), resulting in 
\begin{equation}
	J[k]= \left(\Delta y[k] +\Delta\mathbf{h}^T[k]\boldsymbol{\theta}\right)^2.
	\label{eq_costfun1} 
\end{equation}
Applying the \ac{sgd} method to (\ref{eq_costfun1}) yields the parameter update equation
\begin{equation}
	\boldsymbol{\theta}[k] = \boldsymbol{\theta}[k-1]-\mu\Delta\mathbf{h}[k]\left(\Delta y[k]+\Delta\mathbf{h}^T[k]\boldsymbol{\theta}[k-1]\right), 
	\label{eq_updateeq}
\end{equation}
where $\mu$ denotes the step-size \cite{SGD}. Moreover, the already low computational complexity of the update equation above is further reduced by the large number of zeros in $\Delta\mathbf{h}^T$. The proposed \ac{hee} approach in (\ref{eq_updateeq}) is analyzed with respect to convergence, stability, and practical aspects in the following. 

\subsection{Convergence}
It can be shown that the presented parameter estimation converges on the mean toward the corresponding \ac{mmse} solution \cite{Phd_Thesis_OL}. To prove this, the \ac{mmse} solution is computed first, utilizing the \ac{mse} cost-function
\begin{align}
	J &=\E{\left(\Delta y[k] +\Delta\mathbf{h}^T[k]\boldsymbol{\theta}\right)^2}{},
\end{align}
with $\E{\cdot}{}$ denoting the expected value. Minimizing the equation above yields
\begin{equation}
	\boldsymbol{\theta}_0= -\mathbf{R}_\mathbf{hh}^{-1}\mathbf{r}_{\mathbf{h}y}, 
	\label{eq_Wiener}
\end{equation}
with the autocorrelation matrix $\mathbf{R}_\mathbf{hh}= \E{\Delta\mathbf{h}[k]\Delta\mathbf{h}^T[k]}{}$ and the cross-correlation vector $\mathbf{r}_{\mathbf{h}y} =\E{\Delta\mathbf{h}[k]\Delta y[k]}{}$. Subtracting the \ac{mmse} solution $\boldsymbol{\theta}_0$ from both sides of (\ref{eq_updateeq}), and using $\mathbf{v}[k] = \E{\boldsymbol{\theta}[k]-\boldsymbol{\theta}_0}{}$, results in
\begin{equation}
	\mathbf{v}[k] = \left(\mathbf{I}-\mu\mathbf{R}_\mathbf{hh}\right)\mathbf{v}[k-1].
	\label{eq_conv}
\end{equation}
Assuming that $\mu$ is chosen such that $\| \left(\mathbf{I}-\mu\mathbf{R}_\mathbf{hh}\right)\|_2<1$, it follows that the parameter estimation converges on average toward the \ac{mmse} solution.
\subsection{Stability}
As the convergence proof above requires statistical knowledge of $\Delta\mathbf{h}[k]$, it is more of a theoretical statement. To obtain practical boundaries on the step-size, an alternative way is pursued in the following. Subtracting the \ac{mmse} solution from both sides of (\ref{eq_updateeq}), and denoting $\mathbf{e}[k] = \boldsymbol{\theta}[k]-\boldsymbol{\theta}_0$, allows rewriting (\ref{eq_updateeq}) as
\begin{equation}
	\begin{split}
		\mathbf{e}[k] =&\text{\space} \mathbf{e}[k-1]-\text{\space}\mu\Delta\mathbf{h}[k]\left(\Delta y[k]+\Delta\mathbf{h}^T[k]\left(\mathbf{e}[k-1]+\boldsymbol{\theta}_0\right)\right).
	\end{split}
\end{equation}
With $\boldsymbol{\Lambda}[k] = \left(\mathbf{I}-\mu\Delta\mathbf{h}[k]\Delta\mathbf{h}^T[k]\right)$, and $\Delta\epsilon[k] = \Delta y[k]+\Delta\mathbf{h}^T\boldsymbol{\theta}_0$, the equation above may be written as
\begin{equation}
	\mathbf{e}[k] = \boldsymbol{\Lambda}[k]\mathbf{e}[k-1]-\mu\Delta\mathbf{h}[k]\Delta\epsilon[k].
	\label{eq_err}
\end{equation} 
The term $\Delta \epsilon[k]$ denotes errors not covered by the \ac{mmse} solution, i.\,e. nonidealities in stages not considered for calibration, as well as the \ac{adc} quantization noise. In (\ref{eq_err}) this error is weighted by the step-size $\mu$, by means that although $\Delta \epsilon[k]$ cannot be eliminated, its impact may be reduced by choosing a smaller step-size. Inspecting (\ref{eq_err}) reveals that stability is achieved for $\|\boldsymbol{\Lambda}[k]\|_2<1$. This yields the deterministic step-size boundaries
\begin{equation}
	0<\mu<\frac{2}{\max\limits_k\|\Delta\mathbf{h}[k]\|_2^2}.
	\label{eq_stepsize}
\end{equation}
Fortunately, as shown above $\Delta\mathbf{h}[k]$ originates from the selection vectors according to $\Delta\mathbf{h}[k] = \mathbf{h}_{\alpha x}[k]-\alpha\mathbf{h}_x[k]$. Moreover, if the post correction model is employed in a way, that no entries of $\mathbf{h}_x[k]$ are replaced with $z_k$, as discussed in Sec.\,\ref{sec_postcorr}, the maximum number of ones in these selection vectors corresponds to the number of considered stages $q$. Thus, $\Delta\mathbf{h}[k]$ contains $q$ elements with value 1 and $q$ elements with value $\alpha$. Based on that, the step-size boundaries may be written as
\begin{equation}
	0<\mu<\frac{2}{\left(1+\alpha^2\right)q}.
	\label{eq_stepsize1}
\end{equation}
Nevertheless, if the post correction model is extended by replacing $q$ entries with $z_k$, the values of these $q$ entries depend on the unknown \ac{adc} nonideality. Consequently, for this case, the step-size boundaries cannot be derived a priori as in (\ref{eq_stepsize1}).
\subsection{Practical Aspects}
As discussed at the beginning of this section, the input samples $x_\text{in}[k]$ may originate from a dedicated test \ac{sg} or the \ac{sa} stored in an \ac{sh} block. The latter one has the advantage that the application is not interrupted for calibration. On the downside, storing each sample in an \ac{sh} block and feeding it into the \ac{adc} twice reduces the effective sampling rate of the \ac{adc} by a factor of two. Another drawback might be that the application signal does not necessarily cover the full \ac{adc} transition function, such that only a part of it will be calibrated. Hence, the applicability of using the \ac{hee} approach with the \ac{sa} signal depends on the application. By using a test \ac{sg} and a dedicated calibration time slot, these drawbacks can easily be overcome. To achieve the full potential of the \ac{hee} approach, the authors suggest a two-phase calibration. In the initial phase, after the power-on of the device, an \ac{sg} is used as the test signal source. Within this phase a full scale test signal is injected, such that the entire \ac{adc} transition function is calibrated. After the initial calibration, the signal source is switched to the \ac{sa}. With this second phase, the nonidealities are tracked continuously without repeatedly interrupting the \ac{adc} sampling task. As only slowly changing nonidealities are expected, it is not necessary to use each sample for calibration. Therefore, the impact of the \ac{sh} block regarding \ac{adc} bandwidth can be reduced as well. 
Furthermore, to omit additional digital hardware caused by different estimation algorithms in the two phases, the \ac{als} algorithm \cite{Kaczmarz} may be utilized for the initial calibration. Then both estimators support the same architecture, except for the memory used in the first phase.

\section{Simulation Results}
\label{sec_SIM}

The calibration performance of the \ac{hee} approach is demonstrated with simulations in this section. We consider a 13\,Bit pipelined \ac{adc} with six stages and a sampling rate of 100\,MHz. The stages 1-5 are implemented as 2.5\,Bit stages, while the final flash \ac{adc} resolves 3\,Bits. Regarding nonidealities, the first five stages suffer from gain and \ac{dac} mismatches. All gain and \ac{dac} mismatches were chosen from a uniform \ac{pdf}. This uniform \ac{pdf} was designed such that in the worst case the first stage gain and \ac{dac} mismatches contribute to the overall \ac{adc} nonideality with $\pm25\,\text{LSBs}$ and $\pm15\,\text{LSBs}$, respectively. Note that, mismatches corresponding to less significant stages are weighted appropriately. Additionally, $\alpha=0.5$ is chosen as the scaling factor.\\
\begin{figure}[!t]
	\centering
	\includegraphics{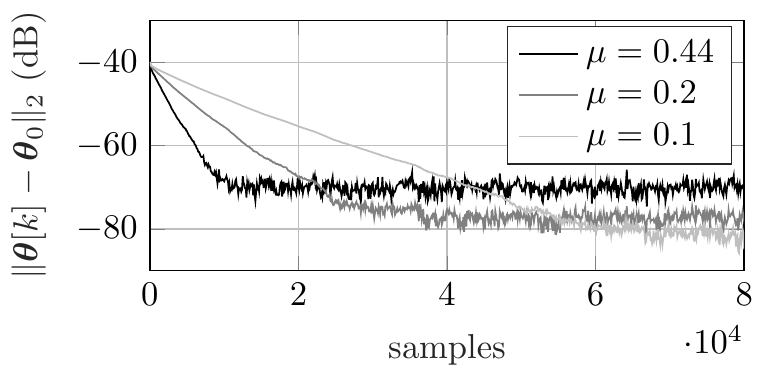}
	\caption{Error norm with respect to different step-sizes.}
	\label{fig_hecConv}
\end{figure}%
The \ac{hee} approach employed in this simulation covers the three most significant stages, which results in 19 parameters to be estimated. A full scale sine wave at $10.77$\,MHz was injected as the test signal. Fig.\,\ref{fig_hecConv} shows the error norm $\|\boldsymbol{\theta}[k]-\boldsymbol{\theta}_0\|_2$ for different step-sizes $\mu$. As can be observed, the error reduces significantly for smaller step-sizes. This reduction results from the fact, that the step-size dependent error term $\Delta\epsilon[k]$ becomes smaller. Nonetheless, smaller step-sizes obviously cause longer calibration times. As a compromise of these two aspects, $\mu = 0.2$ is chosen for all further simulations. With this step-size a full calibration is achieved after $3\cdot 10^4$ samples.%

\begin{figure}[!t]
	\centering
	\includegraphics{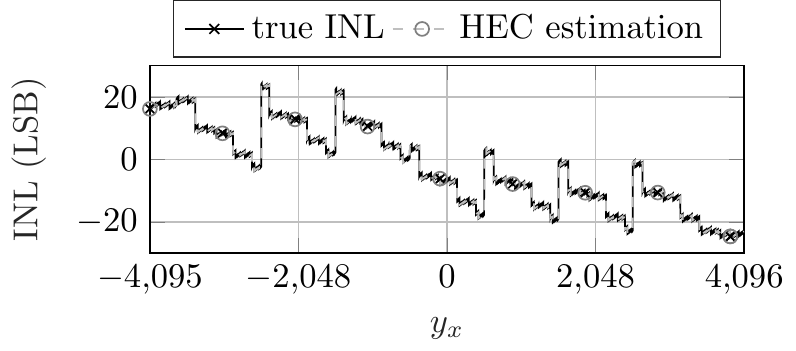}
	\caption{True \ac{inl} and the corresponding \ac{hee} estimation.}
	\label{fig_calPerf}
\end{figure}

\begin{figure}
	\centering
	\includegraphics{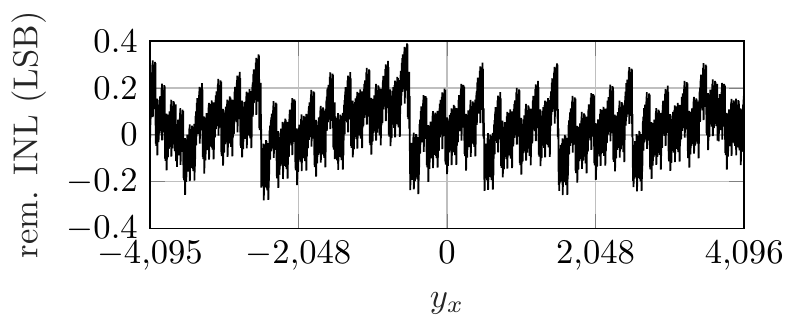}
	\caption{Remaining \ac{inl} after calibration.}
	\label{fig_reminl}
\end{figure}

The calibration performance is verified in Fig.\,\ref{fig_calPerf} and Fig.\,\ref{fig_reminl} regarding the \ac{inl} of the \ac{adc}. The \ac{inl} can be derived by subtracting the input signal $x_\text{in}$ and the quantization error $e_{\text{s},\text{F}}^\text{q}\prod_{i=1}^{n}\frac{ 1}{G_i}$
from (\ref{eq_INL}), and is plotted over the \ac{adc} output. Fig.\,\ref{fig_calPerf} shows that the \ac{inl} identified with the \ac{hee} approach closely matches the true one. The difference in these two \ac{inl} curves is shown in Fig.\,\ref{fig_reminl}, where it can be observed that the \ac{hee} approach achieves a remaining error of less than 0.4\,LSBs. 


\section{Conclusion}
\label{sec_con}
In this work, an equalization-based calibration technique for pipelined \acp{adc}, denoted as the \ac{hee} approach, was proposed. The \ac{hee} approach allows for different test signal sources, such as a test \ac{sg}, or the \ac{sa} signal itself. In the latter case, the \ac{hee} approach corresponds to a background calibration technique. The proposed calibration technique was analyzed with respect to stability and convergence behavior. Ultimately, the achieved calibration performance was verified by simulations.

\section*{Acknowledgment}
This work has been funded by the Linz Center of Mechatronics (LCM) GmbH as part of a K2 project. K2 projects are financed using funding from the Austrian COMET K2 programme. The COMET K2 projects at LCM are supported by the Austrian federal government, the federal state of Upper Austria, the Johannes Kepler University and all of the scientific partners which form part of the COMET K2 consortium.
\newpage


\vfill
%
\end{document}